\documentclass[english]{article}
\usepackage[T1]{fontenc}
\usepackage[latin9]{inputenc}
\usepackage{color}
\usepackage{amstext}
\usepackage{amssymb}
\usepackage{graphicx}
\usepackage{esint}
\PassOptionsToPackage{normalem}{ulem}
\usepackage{ulem}

\makeatletter
\newcommand{\lyxaddress}[1]{
\par {\raggedright #1
\vspace{1.4em}
\noindent\par}
}

\makeatother

\usepackage{babel}
\begin{document}

\title{\textbf{The future of gravitational theories in the era of the gravitational
wave astronomy}}

\author{\textbf{Christian Corda}}
\maketitle

\lyxaddress{\begin{center}
Research Institute for Astronomy and Astrophysics of Maragha (RIAAM),
P.O. Box 55134-441, Maragha, Iran
\par\end{center}}

\begin{center}
\textit{E-mail address:} \textcolor{blue}{cordac.galilei@gmail.com} 
\par\end{center}
\begin{abstract}
We discuss the future of gravitational theories in the framework of
gravitational wave (GW) astronomy after the recent GW detections (the
events GW150914, GW151226, GW170104, GW170814, GW170817 and GW170608).
In particular, a calculation of the frequency and angular dependent
response function that a GW detector would see if massive modes from
$f(R)$ theories or scalar tensor gravity (STG) were present, allowing
for sources incident from any direction on the sky, is shown. In addition,
through separate theoretical results which do not involve the recent
GW detections, we show that $f(R)$ theories of gravity having a third
massless mode are ultimately ruled out while there is still room for
STG having a third (massive or massless) mode and for $f(R)$ theories
of gravity having a third massive mode. 
\end{abstract}
\begin{quotation}
\end{quotation}
\begin{quote}

\emph{To the memory of Ron Drever and Adalberto Giazotto}
\end{quote}

\section{Introduction}

The discovery of GW emissions from the compact binary system with
two neutron stars (NS) PSR1913+16 \cite{key-1} excited interest in
GWs although the first efforts at direct detection started before
that, by involving the design, implementation, and advancement of
extremely sophisticated GW detection technology (see the recent review
\cite{key-31} for the history of GW research). Physicists working
in this field of research need this technology to conduct thorough
investigations of GWs in order to advance science. The main motivation
for searching for GWs is to use them as a probe of the systems that
produce them. The first observation of GWs from a binary black hole
(BH) merger (event GW150914) \cite{key-2}, which occurred in the
100th anniversary of Albert Einstein's prediction of GWs \cite{key-3},
has recently shown that this ambitious challenge has been won. The
event GW150914 represented a cornerstone for science and for gravitational
physics in particular. In fact, this remarkable event equipped scientists
with the means to give definitive proof of the existence of GWs, the
existence of BHs having mass greater than 25 solar masses and the
existence of binary systems of BHs which coalesce in a time less than
the age of the Universe \cite{key-2}. As a consequence of the event
GW150914, the Nobel Prize in Physics 2017 has been awarded to Rainer
Weiss, Barry Barish and Kip Thorne. 

A subsequent analysis of GW150914 constrained the graviton Compton
wavelength of those alternative gravity theories (AGTs) in which the
graviton is massive and placed a lower bound of $10^{13}$ km, corresponding
to a graviton mass $m_{g}\leq1.2\times10^{-22}\frac{eV}{c^{2}}$ \cite{key-4}.
Within their statistical uncertainties, the LIGO Scientific Collaboration
and the Virgo Collaboration have not found evidence of violations
of the general theory of relativity (GTR) in the genuinely strong-field
regime of gravity \cite{key-4}. After the event GW150914, the LIGO
Scientific Collaboration and the Virgo Collaboration announced other
five new GW detections, the events GW151226 \cite{key-34}, GW170104
\cite{key-35}, GW170814 \cite{key-36}, GW170817 \cite{key-37} and
GW170608 \cite{key-38}. All the cited events again arise from binary
BH coalescences with the sole exception ofthe event GW170817, which
represent the first GW detection from a NS merger \cite{key-37}.
After such events, the bound on the graviton mass is even more compelling:
$m_{g}\leq7.7\times10^{-23}\frac{eV}{c^{2}}$ \cite{key-35}. Notice
that this does not mean that the analyses in \cite{key-4,key-35}
show that the true theory of gravity is massless. In fact, the analyses
in \cite{key-4,key-35} have not shown that the graviton mass is zero,
just that it is small. One expects that LIGO and the other GW interferometers
will never show that the mass is exactly 0, they can only place increasingly
precise bounds on its value. On the other hand, the possibility that
AGTs are still alive after the event GW150914 has been emphasized
in \cite{key-5}. In fact, in \cite{key-6} two important questions
have been raised, verbatim: ``\emph{Does gravity really behave as
predicted by Einstein in the vicinity of black holes, where the fields
are very strong? Can dark energy and the acceleration of the Universe
be explained if we modify Einstein's gravity?}'' The current situation
is that ``\emph{We are only just beginning to answer these questions}''
\cite{key-6}. 

Among the various kinds of AGTs, $f(R)$ theories and STG seem to
be the most popular among gravitational physicists because they could
be, in principle, important in order to solve some problem of standard
cosmology like the Dark Matter and Dark Energy problems \cite{key-7,key-8,key-9,key-12}.
These theories attempt to extend the framework of the GTR by modifying
the Lagrangian, with respect to the standard Einstein-Hilbert gravitational
Lagrangian, through the addition of high-order terms in the curvature
invariants (terms like $R^{2}$, $R^{ab}R_{ab}$, $R^{abcd}R_{abcd}$,
$R\Box R$, $R\Box^{k}R$) and/or terms with scalar fields non-minimally
coupled to geometry (terms like $\phi^{2}R$ ) \cite{key-7,key-8,key-9,key-12,key-18}.
In this paper we will focus on these two classes of AGTs. Criticisms
on such theories arises from the fact that lots of them can be excluded
by requirements of cosmology and solar system tests \cite{key-9,key-11,key-15,key-25}.
Thus, one needs the additional assumption that the variation from
the standard GTR must be weak \cite{key-12}. 

For the sake of completeness, as the number of predictions of AGTs
which are highlighted in this paper cannot be tests using the recent
GW detections, it could be useful for the reader to know some ways
in which those predictions might be, in principle, tested in the future.
Following \cite{key-9}, one sees that strong gravity tests are considered,
together with GWs and stellar system tests, the fundamental gravitational
tests for the 21st Century. Those systems for which the simple first
order post-Newtonian approximation (PNA) is no longer appropriate
are called\emph{ strong-field systems} (SFSs) \cite{key-9}. Usually,
SFSs contain strongly relativistic objects, such as NS or BHs, where
the first order PNA breaks down \cite{key-9}. The key point is that
in AGTs the strong-field internal gravity of the bodies should leave
imprints on the orbital motion of the objects \cite{key-9}. SFSs
are also connected with GWs, because GWs can affect the evolution
of the SFS. In fact, as the first order PNA does not contain the effects
of the GW back-reaction, we need a solution of the equations substantially
beyond the first order PNA \cite{key-9}. Concerning stellar system
tests, the discovery of the binary pulsar B1913+16 \cite{key-1} had
importance in the gravitational physics also beyond GWs. Another key
point is indeed the effects of strong relativistic internal gravitational
fields on orbital dynamics \cite{key-9}. Gravitational theory is
today tested with pulsars, including binary and millisecond pulsars
\cite{key-9}. Assuming that both members of the system are NS, the
formulas for the periastron shift, the gravitational redshift/second-order
Doppler shift parameter, the Shapiro delay coefficients, and the rate
of change of orbital period can be obtained \cite{key-9}. On one
hand, the near equality of NS masses in typical double NS binary pulsars
makes bounds obtained not competitive with the Cassini bound \cite{key-25}
because dipole radiation is somewhat suppressed, see \cite{key-9}
and Section 2 of this paper. On the other hand, more promising tests
of dipole radiation arise from a binary pulsar system having dissimilar
objects, such as a white dwarf (WD) or BH companion \cite{key-9}.
An important example is the NS\textendash WD system J1738+0333, which
yields much more stringent bounds, surpassing the Cassini bound \cite{key-9}.
In years to come, the experiments that have been cited will be further
improved and perfected in order to search for new physics beyond Einstein's
GTR.

The main new results of this paper will be shown in next Section and
are the following:
\begin{itemize}
\item We perform a calculation of the frequency and angular dependent response
function that a GW detector would see if massive modes were present,
allowing for sources incident from any direction on the sky. This
will permit, in principle, to discriminate between massless and massive
modes in $f(R)$ theories and STG, while such a discrimination was
not possible in previous GW literature.
\item Through separate theoretical results which do not involve the recent
GW detections, we show that $f(R)$ theories of gravity having a third
massless mode are ultimately ruled out while there is still room for
STG having a third (massive or massless) mode and for $f(R)$ theories
of gravity having a third massive mode. This issue has an important
consequence on the debate on the equivalence or non-equivalence between
$f(R)$ theories an STG \cite{key-7,key-9,key-12,key-32}. 
\end{itemize}

\section{Viability of $f(R)$ theories and scalar tensor gravity through gravitational
waves}

We emphasize that in discussing dipole and monopole radiation we closely
follow the papers \cite{key-14,key-19,key-20}.

In the framework of GWs, the more important difference between the
GTR and the cited two classes of AGTs ($f(R)$ theories and STG) is
the existence, in the latter, of dipole and monopole radiation \cite{key-14,key-19}.
In the GTR, for slowly moving systems, the leading multipole contribution
to gravitational radiation is the quadrupole one, with the result
that the dominant radiation-reaction effects are at order $(\frac{v}{c})^{5}$,
where $v$ is the orbital velocity. The rate, due to quadrupole radiation
in the GTR, at which a binary system loses energy is given by (we
work with $16\pi G=1$, $c=1$ and $\hbar=1$ in the following) \cite{key-14,key-19}

\begin{equation}
\left(\frac{dE}{dt}\right)_{quadrupole}=-\frac{8}{15}\eta^{2}\frac{m^{4}}{r^{4}}(12v^{2}-11\dot{r}^{2}).\label{eq:  Will}
\end{equation}
$\eta$ and $m$ are the reduced mass parameter and total mass, respectively,
given by $\eta=\frac{m_{1}m_{2}}{(m_{1}+m_{2})^{2}}$ , and $m=m_{1}+m_{2}$
. $r,$ $v,$ and $\dot{r}$ represent the orbital separation, relative
orbital velocity, and radial velocity, respectively. 

In $f(R)$ theories and STG, eq. (\ref{eq:  Will}) is modified by
PN corrections to monopole and dipole radiation, and even a cross-term
between dipole and octupole radiation as \cite{key-9} 
\begin{equation}
\frac{dE}{dt}=-\frac{8}{15}\alpha^{3}\eta^{2}\left(\frac{m}{r}\right)^{4}(k_{1}v^{2}-k_{2}\dot{r}^{2}),\label{eq: correction}
\end{equation}
where $\alpha$ is a two-body gravitational interaction parameter
\cite{key-9}, and the parameters $k_{1}$and $k_{2}$ have been calculated
in \cite{key-26}.

The important modification in $f(R)$ theories and STG is the additional
energy loss caused by dipole modes. By analogy with electrodynamics,
dipole radiation is a $(v/c)^{3}$ effect, potentially much stronger
than quadrupole radiation. However, in $f(R)$ theories and STG, the
gravitational \textquotedblleft \emph{dipole moment}\textquotedblright{}
is governed by the difference $S\equiv s_{1}-s_{2}$ between the bodies,
where $s_{i}$ is a measure of the self-gravitational binding energy
per unit rest mass of each body \cite{key-14,key-19}. $s_{i}$ represents
the \textquotedblleft \emph{sensitivity}\textquotedblright{} of the
total mass of the body to variations in the background value of the
Newton constant, which, in this theory, is a function of the scalar
field (an ``effective''scalar field in the case of $f(R)$ theories
\cite{key-11}) \cite{key-14,key-19}: 

\begin{equation}
s_{i}=\left(\frac{\partial(\ln m_{i})}{\partial(\ln G)}\right)_{N}.\label{eq: Will 2}
\end{equation}
\emph{$G$} is the effective Newtonian constant at the star and the
subscript $N$ denotes holding baryon number fixed. 

To first order in $\frac{1}{w}$, where $w$ is the coupling parameter
of the scalar field \cite{key-19}, the energy loss caused by dipole
radiation is given by \cite{key-14,key-19} 
\begin{equation}
\left(\frac{dE}{dt}\right)_{dipole}=-\frac{2}{3}\eta^{2}\frac{m^{4}}{r^{4}}S^{2}.\label{eq:  Will 3}
\end{equation}
In $f(R)$ theories and STG, the sensitivity of a BH is always $s_{BH}=0.5$
\cite{key-14,key-19}, while the sensitivity of a NS varies with the
equation of state and mass. For example, $s_{NS}\approx0.12$ for
a NS of mass order $1.4M_{\circledcirc}$, being $M_{\circledcirc}$
the solar mass \cite{key-14,key-19}. 

Binary BH systems are not at all promising for studying dipole modes
because $s_{BH1}-s_{BH2}=0,$ a consequence of the no-hair theorems
for BHs \cite{key-14,key-19}. BHs indeed radiate away any scalar
field, so that a binary BH system in $f(R)$ theories and STG behaves
as in the GTR. Similarly, binary NS systems are also not effective
testing grounds for dipole radiation \cite{key-14,key-19}. This is
because NS masses tend to cluster around the Chandrasekhar limit of
$1.4M_{\circledcirc}$, and the sensitivity of NSs is not a strong
function of mass for a given equation of state. Thus, in systems like
the binary pulsar, dipole radiation is naturally suppressed by symmetry,
and the bound achievable cannot compete with those from the solar
system \cite{key-14,key-19}. Hence the most promising systems are
mixed: BH-NS, BH-WD or NS-WD. 

The emission of monopole radiation in $f(R)$ theories and STG is
very important in the collapse of quasi-spherical astrophysical objects
because in this case the energy emitted by quadrupole modes can be
neglected \cite{key-14,key-20}. The authors of \cite{key-20} have
shown that, in the formation of a NS, monopole waves interact with
the detectors as well as quadrupole ones. In that case, the field-dependent
coupling strength between matter and the scalar field has been assumed
to be a linear function of the scalar field $\varphi$. In the notation
of this paper such a coupling strength is given by $k^{2}=\frac{16\pi}{|2\omega+3|}$
in eq. (2) of \cite{key-14}. Then \cite{key-20}

\begin{equation}
k^{2}=\alpha_{0}+\beta_{0}(\varphi-\varphi_{0})\label{eq: accoppiamento}
\end{equation}
and the amplitude of the scalar polarization results \cite{key-20}

\begin{equation}
\Phi\propto\frac{\alpha_{0}}{d}\label{eq: ampiezza da supernova}
\end{equation}
where $d$ is the distance of the collapsing NS expressed in meters.

For the following discussion the key point is that STG and $f(R)$
theories have an additional GW polarization which, in general, is
massive with respect to the two standard polarizations of the GTR;
see \cite{key-10,key-11,key-12,key-13,key-14}. As GW detection is
performed in a laboratory environment on Earth, one typically uses
the coordinate system in which space-time is locally flat and the
distance between any two points is given simply by the difference
in their coordinates in the sense of Newtonian physics. This is the
so-called gauge of the local observer \cite{key-10,key-13,key-14,key-16}.
In such a gauge the GWs manifest themselves by exerting tidal forces
on the masses (the mirror and the beam-splitter in the case of an
interferometer) \cite{key-10,key-13,key-14,key-16}. By putting the
beam-splitter in the origin of the coordinate system, the components
of the separation vector are the coordinates of the mirror. The effect
of the GW is to drive the mirror to have oscillations \cite{key-10,key-13,key-14,key-16}.
Let us consider a mirror that has the initial (unperturbed) coordinates
$x_{M0}$, $y_{M0}$ and $z_{M0}$ , where there is a GW propagating
in the $z$ direction. 

In the GTR the GW admits only the standard $+$ and $\times$ polarizations
\cite{key-10,key-16}. We label the respective metric perturbations
as $h_{+}$ and $h_{\times}$. To the first order approximation of
$h_{+}$ and $h_{\times}$ the motion of the mirror due to the GW
is \cite{key-10,key-16} 
\begin{equation}
\begin{array}{c}
x_{M}(t)=x_{M0}+\frac{1}{2}[x_{M0}h_{+}(t)-y_{M0}h_{\times}(t)]\\
\\
y_{M}(t)=y_{M0}-\frac{1}{2}[y_{M0}h_{+}(t)+x_{M0}h_{\times}(t)]\\
\\
z_{M}(t)=z_{M0}.
\end{array}\label{eq: traditional GTR}
\end{equation}
STG can be have a third additional mode that is massless \cite{key-10,key-12,key-14}.
In this case, calling $h_{\Phi}$ the metric perturbation due to the
additional GW polarization, to the first order approximation of $h_{+}$,
$h_{\times}$ and $h_{\Phi}$ the motion of the mirror due to the
GW is \cite{key-10,key-14}

\begin{equation}
\begin{array}{c}
x_{M}(t)=x_{M0}+\frac{1}{2}[x_{M0}h_{+}(t)-y_{M0}h_{\times}(t)]+\frac{1}{2}x_{M0}h_{\Phi}(t)\\
\\
y_{M}(t)=y_{M0}-\frac{1}{2}[y_{M0}h_{+}(t)+x_{M0}h_{\times}(t)]+\frac{1}{2}y_{M0}h_{\Phi}(t)\\
\\
z_{M}(t)=z_{M0}.
\end{array}\label{eq: massless STG}
\end{equation}
$f(R)$ theories have a third additional mode which is generally massive
\cite{key-11,key-12,key-13,key-14}. The cases of STG and $f(R)$
theories having a third massive additional mode are totally equivalent
\cite{key-11,key-12,key-13,key-14}. This is not surprising because
it is well known that there is a more general conformal equivalence
between $f(R)$ theories and STG \cite{key-7,key-9,key-12,key-32}.
Again, we call $h_{\Phi}$ the metric perturbation due to the additional
GW polarization. To the first order approximation of $h_{+}$, $h_{\times}$
and $h_{\Phi}$ the motion of the mirror due to the GW in STG and
$f(R)$ theories having a third massive additional mode is \cite{key-12,key-13,key-14}
\begin{equation}
\begin{array}{c}
x_{M}(t)=x_{M0}+\frac{1}{2}[x_{M0}h_{+}(t)-y_{M0}h_{\times}(t)]+\frac{1}{2}x_{M0}h_{\Phi}(t)\\
\\
y_{M}(t)=y_{M0}-\frac{1}{2}[y_{M0}h_{+}(t)+x_{M0}h_{\times}(t)]+\frac{1}{2}y_{M0}h_{\Phi}(t)\\
\\
z_{M}(t)=z_{M0}+\frac{1}{2}z_{M0}\frac{m^{2}}{\omega^{2}}h_{\Phi}(t),
\end{array}\label{eq: massive polarization}
\end{equation}
where $m$ and $\omega$ are the mass and the frequency of the GW's
third massive mode, which is interpreted in terms of a wave packet
\cite{key-12,key-13,key-14}. We also recall that the relation between
the mass and the frequency of the wave packet is given by \cite{key-11,key-13,key-14}
\begin{equation}
m=\sqrt{(1-v_{G}^{2})}\omega,\label{eq: relazione massa-frequenza}
\end{equation}
where $v_{G}$ is the group-velocity of the wave-packet. Inserting
eq. (\ref{eq: relazione massa-frequenza}) in the third of eqs. (\ref{eq: massive polarization})
one gets 
\begin{equation}
\begin{array}{c}
x_{M}(t)=x_{M0}+\frac{1}{2}[x_{M0}h_{+}(t)-y_{M0}h_{\times}(t)]+\frac{1}{2}x_{M0}h_{\Phi}(t)\\
\\
y_{M}(t)=y_{M0}-\frac{1}{2}[y_{M0}h_{+}(t)+x_{M0}h_{\times}(t)]+\frac{1}{2}y_{M0}h_{\Phi}(t)\\
\\
z_{M}(t)=z_{M0}+\frac{(1-v_{G}^{2})}{2}z_{M0}h_{\Phi}(t).
\end{array}\label{eq: massive polarization 2}
\end{equation}
The presence of the little mass $m$ implies that the speed of the
third massive mode is less than the speed of light; this generates
the longitudinal component and drives the mirror oscillations of the
$z$ direction \cite{key-11,key-13,key-14}, which is shown by the
third of eqs. (\ref{eq: massive polarization}). 

Now, we perform a calculation of the signal that a GW detector would
see if massive modes were present, allowing for sources incident from
any direction on the sky. For the third massive additional mode the
equation for geodesic deviation gives \cite{key-13} 
\begin{equation}
\begin{array}{c}
\widetilde{R}_{010}^{1}=-\frac{1}{2}\ddot{h}_{\Phi}\\
\\
\widetilde{R}_{020}^{2}=-\frac{1}{2}\ddot{h}_{\Phi}\\
\\
\widetilde{R}_{030}^{3}=\frac{1}{2}m^{2}h_{\Phi},
\end{array}\label{eq: componenti riemann}
\end{equation}
where the $\widetilde{R}_{0i0}^{i}$ are the non zero components of
the linearized Riemann tensor \cite{key-13}. Now, let us consider
a GW propagating in an arbitrary direction $\widehat{n}$ with the
arm of the interferometer in the $\hat{u}$ and $\hat{v}$ directions,
see Figure 1. Eq. (\ref{eq: componenti riemann}) can be rewritten
in compact form as 
\begin{equation}
\begin{array}{c}
\widetilde{R}_{0j0}^{i}=\frac{1}{2}\left(\begin{array}{ccc}
-\ddot{h}_{\Phi} & 0 & 0\\
0 & -\ddot{h}_{\Phi} & 0\\
0 & 0 & m^{2}h_{\Phi}
\end{array}\right)=\\
\\
=\frac{1}{2}\ddot{h}_{\Phi}\left(\delta_{ij}-\widehat{n}_{i}\widehat{n}_{j}\right)x_{j}-\frac{1}{2}m^{2}h_{\Phi}\left(\widehat{n}_{i}\widehat{n}_{j}\right)x_{j}.
\end{array}\label{eq: compatta}
\end{equation}
It is possible to associate to the interferometer a \textit{polarization
tensor} defined by \cite{key-29}
\begin{equation}
d^{ij}\equiv\frac{1}{2}(\hat{v}^{i}\hat{v}^{j}-\hat{u}^{i}\hat{u}^{j}).\label{eq: definizione D}
\end{equation}
 
\begin{figure}
\includegraphics[scale=0.7]{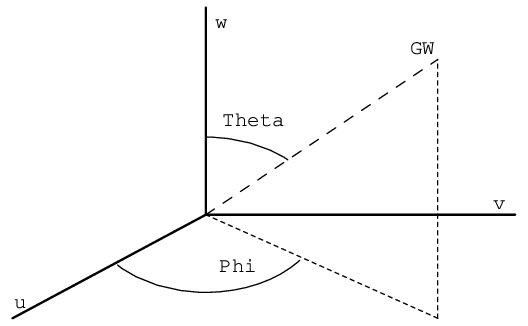}

\caption{a GW propagating in an arbitrary direction, adapted from ref. \cite{key-10}}
\end{figure}
In that case, the signal induced by a generic GW polarization is the
phase shift which is given by \cite{key-29,key-30}

\begin{equation}
s(t)\sim d^{ij}\widetilde{R}_{i0j0}.\label{eq: legame onda-output}
\end{equation}
Using eqs. (\ref{eq: compatta}) and (\ref{eq: definizione D}), one
gets

\begin{equation}
s(t)\sim-\sin^{2}\theta\cos2\phi.\label{eq: legame onda-output 2}
\end{equation}
The angular dependence (\ref{eq: legame onda-output 2}) is different
from the two well known ones arising from the standard tensor modes
of the GTR which are $(1+\cos^{2}\theta)\cos2\phi$ for the $+$ polarization
and $-\cos\theta\sin2\vartheta$ for the $\times$ polarization respectively
\cite{key-30}. Now, let us see what happens for the third additional
massless mode in STG. In that case, eq. (\ref{eq: componenti riemann})
reduces to 
\begin{equation}
\begin{array}{c}
\widetilde{R}_{010}^{1}=-\frac{1}{2}\ddot{h}_{\Phi}\\
\\
\widetilde{R}_{020}^{2}=-\frac{1}{2}\ddot{h}_{\Phi},
\end{array}\label{eq: componenti riemann 2}
\end{equation}
which can be rewritten in compact form as 
\begin{equation}
\begin{array}{c}
\widetilde{R}_{0j0}^{i}=\frac{1}{2}=\left(\begin{array}{cc}
-\ddot{h}_{\Phi} & 0\\
0 & -\ddot{h}_{\Phi}
\end{array}\right)\\
\\
=\frac{1}{2}\ddot{h}_{\Phi}\left(\delta_{ij}-\widehat{n}_{i}\widehat{n}_{j}\right)x_{j}.
\end{array}\label{eq: compatta 2}
\end{equation}
Then, using eqs. (\ref{eq: compatta 2}) and (\ref{eq: definizione D}),
one gets again

\begin{equation}
s(t)\sim-\sin^{2}\theta\cos2\phi,\label{eq: legame onda-output 3}
\end{equation}
which is the same result of eq. (\ref{eq: legame onda-output 2}).
Thus, by using this approach one cannot discriminate between massless
and massive modes. On the other hand, the results of eqs. (\ref{eq: legame onda-output 2})
and (\ref{eq: legame onda-output 3}) are well known, \cite{key-9,key-30}.
Now, in order to discriminate between massless and massive modes we
will compute the frequency and angular dependent response function
of a GW interferometric detector for massive modes. In fact, the angular
dependences (\ref{eq: legame onda-output 2}) and (\ref{eq: legame onda-output 3})
have been computed with the implicit, standard assumption that the
GW-wavelength is much larger than the distance between the test masses,
which are the two mirrors and the beam-splitter for interferometers
like LIGO \cite{key-10,key-30}. This low frequency approximation
does not permit to discriminate between massless and massive modes
because the angular dependence is the same in both of the cases. We
will go beyond the low frequency approximation in the next analysis.
To compute the response function of the interferometer to a massive
mode from arbitrary propagating directions we must perform a spatial
rotation of the coordinate system as 

\begin{equation}
\begin{array}{ccc}
u & = & -x\cos\theta\cos\phi+y\sin\phi+z\sin\theta\cos\phi\\
\\
v & = & -x\cos\theta\sin\phi-y\cos\phi+z\sin\theta\sin\phi\\
\\
w & = & x\sin\theta+z\cos\theta,
\end{array}\label{eq: rotazione}
\end{equation}
or, in terms of the $x,y,z$ frame:

\begin{equation}
\begin{array}{ccc}
x & = & -u\cos\theta\cos\phi-v\cos\theta\sin\phi+w\sin\theta\\
\\
y & = & u\sin\phi-v\cos\phi\\
\\
z & = & u\sin\theta\cos\phi+v\sin\theta\sin\phi+w\cos\theta.
\end{array}\label{eq: rotazione 2}
\end{equation}
The test masses are the beam splitter and the mirror of the interferometer,
and we will suppose that the beam splitter is located in the origin
of the coordinate system. Hence, Eqs. (\ref{eq: massive polarization 2})
represent the motion of the mirror like it is due to the massive mode
of the GW. The mirror of Eqs. (\ref{eq: massive polarization 2})
is situated in the $u$ direction. Thus, using Eqs. (\ref{eq: massive polarization 2}),
(\ref{eq: rotazione}) and (\ref{eq: rotazione 2}) the $u$ coordinate
of the mirror is given by

\begin{equation}
\begin{array}{c}
u_{M}=-\left(x_{M0}+\frac{1}{2}x_{M0}h_{\Phi}(t)\right)\left(\cos\theta\cos\phi\right)\\
\\
+\left(y_{M0}+\frac{1}{2}y_{M0}h_{\Phi}(t)\right)\sin\phi\\
\\
+\left(z_{M0}+\frac{(1-v_{G}^{2})}{2}z_{M0}h_{\Phi}(t)\right)\left(z\sin\theta\cos\phi\right).
\end{array}\label{eq: du}
\end{equation}
In the same way, the $v$ coordinate of the mirror is given by 
\begin{equation}
\begin{array}{c}
v_{M}=-\left(x_{M0}+\frac{1}{2}x_{M0}h_{\Phi}(t)\right)\left(\cos\theta\sin\phi\right)\\
\\
-\left(y_{M0}+\frac{1}{2}y_{M0}h_{\Phi}(t)\right)\cos\phi\\
\\
+\left(z_{M0}+\frac{(1-v_{G}^{2})}{2}z_{M0}h_{\Phi}(t)\right)\left(z\sin\theta\sin\phi\right).
\end{array}\label{eq: dv}
\end{equation}
Following \cite{key-10,key-12,key-13,key-33}, a good way to analyse
variations in the proper distance (time) is by means of ``bouncing
photons''. A photon can be launched from the interferometer's beam-splitter
to be bounced back by the mirror. The ``bouncing photons analysis''
was created in \cite{key-33}. Actually, it has strongly generalized
to angular dependences and scalar waves in \cite{key-10,key-12,key-13}
but this is the first time that such an analysis is performed in order
to compute the frequency and angular dependent response function for
massive modes. We will consider a photon propagating in the $u$ axis.
The analysis is similar for a photon propagating in the $v$ axis.
By using eq. (\ref{eq: du}), the unperturbed coordinates for the
beam-splitter and the mirror are $u_{b}=0$ and $u_{m}=L$, where
$L=\sqrt{x_{M0}^{2}+y_{M0}^{2}+z_{M0}^{2}}$ is the length of the
interferometer arms. Then, the unperturbed propagation time between
the two masses is

\begin{equation}
T=L.\label{eq: tempo imperturbato}
\end{equation}
From eq. (\ref{eq: du}), one gets the displacements of the two masses
under the influence of the massive mode of the GW as

\begin{equation}
\delta u_{BS}(t)=0\label{eq: spostamento beam-splitter}
\end{equation}
and

\begin{equation}
\delta u_{M}(t)=\frac{1}{2}Ah_{\Phi}(t+L\sin\theta\cos\phi),\label{eq: spostamento mirror}
\end{equation}
where 
\begin{equation}
A\equiv-x_{M0}\cos\theta\cos\phi+y_{M0}\sin\phi+z_{M0}\sin\theta\cos\phi.\label{eq: A}
\end{equation}
Therefore, the relative displacement in the $u$ direction, which
is defined by

\begin{equation}
\delta L(t)=\delta u_{M}(t)-\delta u_{BS}(t)\label{eq: spostamento relativo}
\end{equation}
gives a ``signal'' in the $u$ direction

\begin{equation}
\frac{\delta T(t)}{T}|_{u}=\frac{\delta L(t)}{L}=\frac{1}{2}\frac{A}{L}h_{\Phi}(t+L\sin\theta\cos\phi).\label{eq: strain massiccio}
\end{equation}
But one sees that for a large separation between the test masses (in
the case of LIGO the distance between the beam-splitter and the mirror
is four kilometres), the definition (\ref{eq: spostamento relativo})
for relative displacements becomes unphysical because the two test
masses are taken at the same time and therefore cannot be in a casual
connection \cite{key-10,key-13,key-33}. Thus, the correct definitions
for the bouncing photon are

\begin{equation}
\delta L_{1}(t)=\delta u_{M}(t)-\delta u_{BS}(t-T_{1})\label{eq: corretto spostamento B.S. e M.}
\end{equation}
and

\begin{equation}
\delta L_{2}(t)=\delta u_{M}(t-T_{2})-\delta u_{BS}(t),\label{eq: corretto spostamento B.S. e M. 2}
\end{equation}
where $T_{1}$ and $T_{2}$ are the photon propagation times for the
forward and return trip correspondingly. Through the new definitions,
the displacement of one test mass is compared with the displacement
of the other at a later time to allow for finite delay from the light
propagation \cite{key-10,key-13,key-33}. The propagation times $T_{1}$
and $T_{2}$ in Eqs. (\ref{eq: corretto spostamento B.S. e M.}) and
(\ref{eq: corretto spostamento B.S. e M. 2}) can be replaced with
the nominal value $T$ because the test mass displacements are already
first order in $h_{\Phi}$ \cite{key-10,key-13,key-33}. In this way,
the total change in the distance between the beam splitter and the
mirror in one round-trip of the photon is

\begin{equation}
\delta L_{r.t.}(t)=\delta L_{1}(t-T)+\delta L_{2}(t)=2\delta u_{m}(t-T)-\delta u_{BS}(t)-\delta u_{BS}(t-2T),\label{eq: variazione distanza propria}
\end{equation}
and in terms of the amplitude of the massive GW mode:

\begin{equation}
\delta L_{r.t.}(t)=Ah_{\Phi}(t+L\sin\theta\cos\phi-L).\label{eq: variazione distanza propria 2}
\end{equation}
The change in distance (\ref{eq: variazione distanza propria 2})
leads to changes in the round-trip time for photons propagating between
the beam-splitter and the mirror in the $u$ direction:

\begin{equation}
\frac{\delta_{1}T(t)}{T}|_{u}=\frac{A}{L}h_{\Phi}(t+L\sin\theta\cos\phi-L).\label{eq: variazione tempo proprio 1}
\end{equation}
One observes that in the last calculation, which concerns the variations
in the photon round-trip time which come from the motion of the test
masses inducted by the massive GW mode, it has been implicitly assumed
that the propagation of the photon between the beam-splitter and the
mirror of the interferometer is uniform as if it were moving in a
flat space-time. But the presence of the tidal forces indicates that
the space-time is curved instead. As a result, one must analyse one
more effect after the first discussed, that requires spacial separation
\cite{key-10,key-13,key-33}. From equation (\ref{eq: spostamento mirror})
the tidal acceleration of a test mass caused by the massive GW mode
in the $u$ direction is 
\begin{equation}
\ddot{u}(t+u\sin\theta\cos\phi)=\frac{1}{2}A\ddot{h}_{\Phi}(t+u\sin\theta\cos\phi).\label{eq: acc}
\end{equation}
This is equivalent to the presence of a gravitational potential \cite{key-10,key-13,key-33}:

\begin{equation}
V(u,t)=-\frac{1}{2}A\int_{0}^{u}\ddot{h}_{\Phi}(t+l\sin\theta\cos\phi)dl,\label{eq:potenziale in gauge Lorentziana}
\end{equation}
generating the tidal forces. Thus, and the motion of the test mass
is governed by the Newtonian equation \cite{key-10,key-13,key-33}

\begin{equation}
\ddot{\overrightarrow{r}}=-\bigtriangledown V.\label{eq: Newtoniana}
\end{equation}
Now, we can discuss the second effect. Let us consider the interval
for photons propagating along the $u$ -axis
\begin{equation}
ds^{2}=g_{00}dt^{2}+du^{2}.\label{eq: metrica osservatore locale}
\end{equation}
The condition for a null trajectory ($ds=0$) gives the coordinate
velocity of the photons \cite{key-10,key-13,key-33}

\begin{equation}
v_{p}^{2}\equiv(\frac{du}{dt})^{2}=1+2V(t,u),\label{eq: velocita' fotone in gauge locale}
\end{equation}
which to first order in $h_{\Phi}$ is approximated by

\begin{equation}
v_{p}\approx\pm[1+V(t,u)],\label{eq: velocita fotone in gauge locale 2}
\end{equation}
with $+$ and $-$ for the forward and return trip respectively. By
knowing the coordinate velocity of the photon, the propagation time
for its travelling between the beam-splitter and the mirror can be
defined as \cite{key-10,key-13,key-33}

\begin{equation}
T_{1}(t)=\int_{u_{BS}(t-T_{1})}^{u_{M}(t)}\frac{du}{v_{p}}\label{eq:  tempo di propagazione andata gauge locale}
\end{equation}
and

\begin{equation}
T_{2}(t)=\int_{u_{M}(t-T_{2})}^{u_{BS}(t)}\frac{(-du)}{v_{p}}.\label{eq:  tempo di propagazione ritorno gauge locale}
\end{equation}
The calculations of these integrals would be complicated because the
$u_{M}$ boundaries of them are changing with time \cite{key-10,key-13,key-33}

\begin{equation}
u_{BS}(t)=0\label{eq: variazione b.s. in gauge locale}
\end{equation}
and

\begin{equation}
u_{M}(t)=L+\delta u_{M}(t).\label{eq: variazione specchio nin gauge locale}
\end{equation}
But, to first order in $h_{\Phi}$, these contributions can be approximated
by $\delta L_{1}(t)$ and $\delta L_{2}(t)$ (see Eqs. (\ref{eq: corretto spostamento B.S. e M.})
and (\ref{eq: corretto spostamento B.S. e M. 2})) \cite{key-10,key-13,key-33}.
Hence, the combined effect of the varying boundaries is given by $\delta_{1}T(t)$
in eq. (\ref{eq: variazione tempo proprio 1}). Therefore, one needs
to compute only the times for photon propagation between the fixed
boundaries, i.e $0$ and $L$. Such propagation times are denoted
with $\Delta T_{1,2}$ to distinguish from $T_{1,2}$. In the forward
trip, the propagation time between the fixed limits is

\begin{equation}
\Delta T_{1}(t)=\int_{0}^{L}\frac{du}{v(t',u)}\approx L-\int_{0}^{L}V(t',u)du,\label{eq:  tempo di propagazione andata  in gauge locale}
\end{equation}
where $t'$ is the delay time (i.e. $t$ is the time at which the
photon arrives in the position $L$, so $L-u=t-t'$ \cite{key-10,key-13,key-33})
which corresponds to the unperturbed photon trajectory: $t'=t-(L-u)$.
Similarly, the propagation time in the return trip is

\begin{equation}
\Delta T_{2}(t)=L-\int_{L}^{0}V(t',u)du,\label{eq:  tempo di propagazione andata  in gauge locale 2}
\end{equation}
where now the delay time is given by $t'=t-u$. The sum of $\Delta T_{1}(t-T)$
and $\Delta T_{2}(t)$ gives the round-trip time for photons travelling
between the fixed boundaries. Then, the deviation of this round-trip
time (distance) from its unperturbed value $2T$ is
\begin{equation}
\begin{array}{c}
\delta_{2}T(t)=-\int_{0}^{L}[V(t-2L+u,u)du\\
\\
-\int_{L}^{0}V(t-u,u)]du,
\end{array}\label{eq: variazione tempo proprio 2}
\end{equation}
and, using Eq. (\ref{eq:potenziale in gauge Lorentziana}), it is

\begin{equation}
\begin{array}{c}
\delta_{2}T(t)=\frac{1}{2}A\int_{0}^{L}[\int_{0}^{u}\ddot{h}_{\Phi}(t-2T+l(1+\sin\theta\cos\phi))dl\\
\\
-\int_{0}^{u}\ddot{h}_{\Phi}(t-l(1-\sin\theta\cos\phi)dl]du.
\end{array}\label{eq: variazione tempo proprio 2 rispetto h}
\end{equation}
Thus, the total round-trip proper distance in presence of the\emph{
}massive GW mode is:

\begin{equation}
T_{t}=2T+\delta_{1}T+\delta_{2}T,\label{eq: round-trip  totale in gauge locale}
\end{equation}
and
\begin{equation}
\delta T_{u}=T_{t}-2T=\delta_{1}T+\delta_{2}T\label{eq:variaz round-trip totale in gauge locale}
\end{equation}
is the total variation of the proper time (distance) for the round-trip
of the photon in presence of the\emph{ }massive GW mode in the $u$
direction. By using Eqs. (\ref{eq: variazione tempo proprio 1}),
(\ref{eq: variazione tempo proprio 2 rispetto h}) and the Fourier
transform of $h_{\Phi}$ defined by 
\begin{equation}
\tilde{h}_{\Phi}(\omega)=\int_{-\infty}^{\infty}dt\,h_{\Phi}\exp(i\omega t),\label{eq: trasformata di fourier}
\end{equation}
the quantity (\ref{eq:variaz round-trip totale in gauge locale})
can be computed in the frequency domain by using the derivation and
translation theorems of the Fourier transform as 

\begin{equation}
\tilde{\delta}T_{u}(\omega)=\tilde{\delta}_{1}T(\omega)+\tilde{\delta}_{2}T(\omega)\label{eq:variaz round-trip totale in gauge locale 2}
\end{equation}
where

\begin{equation}
\tilde{\delta}_{1}T(\omega)=-i\omega\exp[i\omega T(1-\sin\theta\cos\phi)]A\,\tilde{h}_{\Phi}(\omega)\label{eq: dt 1 omega}
\end{equation}
and 

\begin{equation}
\begin{array}{c}
\tilde{\delta}_{2}T(\omega)=-\frac{A\omega^{2}}{2}\exp\left(2i\omega T\right)\left[\frac{T}{i\omega\left(1+\sin\theta\cos\phi\right)}+\frac{\exp[i\omega T(1+\sin\theta\cos\phi)]-1}{\omega^{2}\left(1+\sin\theta\cos\phi\right)^{2}}\right]\,\tilde{h}_{\Phi}(\omega)\\
\\
+\frac{A\omega^{2}}{2}\left[\frac{T}{i\omega\left(1-\sin\theta\cos\phi\right)}+\frac{\exp[i\omega T(1-\sin\theta\cos\phi)]-1}{\omega^{2}\left(1-\sin\theta\cos\phi\right)^{2}}\right]\,\tilde{h}_{\Phi}(\omega).
\end{array}\label{eq: dt 2 omega}
\end{equation}
In this way, one finds the response function of the $u$ arm of the
interferometer to the massive GW mode as

\begin{equation}
\begin{array}{c}
H_{u}^{massive}(\omega)\equiv\frac{\tilde{\delta}T_{u}(\omega)}{T\,\tilde{h}_{\Phi}(\omega)}=\\
\\
=-i\omega\exp[i\omega T(1-\sin\theta\cos\phi)]\frac{A}{T}\\
\\
-\frac{A\omega^{2}}{2T}\exp\left(2i\omega T\right)\left[\frac{T}{i\omega\left(1+\sin\theta\cos\phi\right)}+\frac{\exp[i\omega T(1+\sin\theta\cos\phi)]-1}{\omega^{2}\left(1+\sin\theta\cos\phi\right)^{2}}\right]\\
\\
+\frac{A\omega^{2}}{2T}\left[\frac{T}{i\omega\left(1-\sin\theta\cos\phi\right)}+\frac{\exp[i\omega T(1-\sin\theta\cos\phi)]-1}{\omega^{2}\left(1-\sin\theta\cos\phi\right)^{2}}\right].
\end{array}\label{eq: risposta u}
\end{equation}
The computation for the $v$ arm is parallel to the one above. With
the same way of thinking of previous analysis, defining 
\begin{equation}
B\equiv-x_{M0}\cos\theta\sin\phi-y_{M0}\cos\phi+z_{M0}\sin\theta\sin\phi,\label{eq: B}
\end{equation}
a straightforward similar computation permits to find the response
function of the $v$ arm of the interferometer to the massive GW mode
as 
\begin{equation}
\begin{array}{c}
H_{v}^{massive}(\omega)\equiv\frac{\tilde{\delta}T_{v}(\omega)}{T\,\tilde{h}_{\Phi}(\omega)}=\\
\\
=-i\omega\exp[i\omega T(1-\sin\theta\sin\phi)]\frac{B}{T}\\
\\
-\frac{B\omega^{2}}{2T}\exp\left(2i\omega T\right)\left[\frac{T}{i\omega\left(1+\sin\theta\sin\phi\right)}+\frac{\exp[i\omega T(1+\sin\theta\sin\phi)]-1}{\omega^{2}\left(1+\sin\theta\sin\phi\right)^{2}}\right]\\
\\
+\frac{B\omega^{2}}{2T}\left[\frac{T}{i\omega\left(1-\sin\theta\sin\phi\right)}+\frac{\exp[i\omega T(1-\sin\theta\sin\phi)]-1}{\omega^{2}\left(1-\sin\theta\sin\phi\right)^{2}}\right].
\end{array}\label{eq: risposta v}
\end{equation}
The total response function to the massive GW mode is given by the
difference of the two response function of the two arms:
\begin{equation}
H_{tot}^{massive}(\omega)\equiv H_{u}^{m}(\omega)-H_{v}^{m}(\omega),\label{eq: risposta totalissima}
\end{equation}
and using Eqs. (\ref{eq: risposta u}) and (\ref{eq: risposta v})
one gets
\begin{equation}
\begin{array}{c}
H_{tot}^{massive}(\omega)=\frac{\tilde{\delta}T_{tot}(\omega)}{T\tilde{h}_{\Phi}(\omega)}=\\
\\
=-i\omega\exp[i\omega T(1-\sin\theta\cos\phi)]\frac{A}{T}-i\omega\exp[i\omega T(1-\sin\theta\sin\phi)]\frac{B}{T}\\
\\
-\frac{A\omega^{2}}{2T}\exp\left(2i\omega T\right)\left[\frac{T}{i\omega\left(1+\sin\theta\cos\phi\right)}+\frac{\exp[i\omega T(1+\sin\theta\cos\phi)]-1}{\omega^{2}\left(1+\sin\theta\cos\phi\right)^{2}}\right]\\
\\
-\frac{B\omega^{2}}{2T}\exp\left(2i\omega T\right)\left[\frac{T}{i\omega\left(1+\sin\theta\sin\phi\right)}+\frac{\exp[i\omega T(1+\sin\theta\sin\phi)]-1}{\omega^{2}\left(1+\sin\theta\sin\phi\right)^{2}}\right]\\
\\
+\frac{A\omega^{2}}{2T}\left[\frac{T}{i\omega\left(1-\sin\theta\cos\phi\right)}+\frac{\exp[i\omega T(1-\sin\theta\cos\phi)]-1}{\omega^{2}\left(1-\sin\theta\cos\phi\right)^{2}}\right]\\
\\
+\frac{B\omega^{2}}{2T}\left[\frac{T}{i\omega\left(1-\sin\theta\sin\phi\right)}+\frac{\exp[i\omega T(1-\sin\theta\sin\phi)]-1}{\omega^{2}\left(1-\sin\theta\sin\phi\right)^{2}}\right].
\end{array}\label{eq: risposta totale 2}
\end{equation}
On the other hand, the frequency and angular dependent response function
of a GW interferometric detector for the third massless mode in STG
is well known (see for example \cite{key-12}) and is given by \cite{key-12}
\begin{equation}
\begin{array}{c}
H_{tot}^{massless}(\omega)=\frac{\sin\theta}{2i\omega L}\{\cos\phi[1+\exp(2i\omega L)-2\exp i\omega L(1+\sin\theta\cos\phi)]+\\
\\
-\sin\phi[1+\exp(2i\omega L)-2\exp i\omega L(1+\sin\theta\sin\phi)]\},
\end{array}\label{eq: risposta totale Virgo}
\end{equation}
which is different from eq. (\ref{eq: risposta totale 2}). Thus,
in principle, the frequency and angular dependent response functions
(\ref{eq: risposta totale 2}) and (\ref{eq: risposta totale Virgo})
can be used to discriminate between massless and massive modes in
STG and massive $f(R)$ gravity. Now, one notes that in the low frequency
approximation, that is when $\omega\rightarrow0,$ one gets 
\begin{equation}
H_{tot}^{massless}(\omega)\approx H_{tot}^{massive}(\omega)\approx-\sin^{2}\theta\cos2\phi.\label{eq: basse frequenze}
\end{equation}
Then, one finds again the angular dependences (\ref{eq: legame onda-output 2})
and (\ref{eq: legame onda-output 3}). Thus, the angular dependences
(\ref{eq: legame onda-output 2}) and (\ref{eq: legame onda-output 3})
are sufficient to discriminate between the GTR on one hand and STG
and $f(R)$ gravity on the other hand, but they are not sufficient
to discriminate between massless and massive modes. In order to discriminate
between massless and massive modes one must look at higher frequencies
by using the frequency and angular dependent response functions (\ref{eq: risposta totale 2})
and (\ref{eq: risposta totale Virgo}). This is, in principle, possible,
because the frequency-range for earth based gravitational antennas
is the interval $10Hz\leq f\leq10KHz$ \cite{key-10}. Thus, \uline{eq.
(59) is very important and represents the main result of this paper}.
It is indeed a completely new and original result which can, in principle,
be used to find massive modes arising from STG and $f(R)$ gravity
in the motion of the interferometer's test masses. Instead, such a
discrimination between massive and massless modes was \uline{not}
possible in previous GW literature.

The recent detections imply that the graviton mass must be $m_{g}\leq7.7\times10^{-23}\frac{eV}{c^{2}}$
(in standard units) \cite{key-35}. An important point is that the
graviton mass $m_{g}$ has not to be confused with the quantity $m$
in eq. (\ref{eq: massive polarization}). In fact, the LIGO constraint
in \cite{key-35} is not on the extra polarization mode, which vanishes
in the GTR limit, but on the tensor modes. A further clarification
is needed in order to avoid confusion. Eqs. (\ref{eq: massless STG})
and (\ref{eq: massive polarization}) could give the reader an incorrect
impression of how the works \cite{key-4,key-35} placed constraints
on the graviton mass using the recent GW detections. In fact, eqs.
(\ref{eq: massless STG}) and (\ref{eq: massive polarization}) show
a difference in the $z-$motion of the mirrors when the AGT has a
third propagation mode. However, LIGO has no sensitivity to motion
in the $z-$direction. Of course, if the GW does not come from overhead
the motion will be in the sensitive direction. In any case, even in
the massless case there is a change in the motion of the mirrors due
to the $h_{\Phi}$ term. To detect that in the data, it is necessary
to measure the polarization of the GWs, but LIGO is not very well
set up to do that, since the two detectors are almost aligned. The
addition of Virgo \cite{key-23,key-24} will make that measurement,
in principle, possible, see also the final discussion of this paper
on the realization of a network of interferometers. Moreover, a key
point is that, at the present time, the constraint on the mass is
made indirectly \cite{key-4,key-35}. It comes from the lack of observed
dispersion in the GW signal - an inspiraling binary radiates at different
frequencies as the orbit decays. These different frequencies propagate
at different speeds in massive gravity theories. Thus, the observed
signal suffers dispersion \cite{key-4,key-35}. As we stressed above,
this is in the tensor part of the signal. But the tensor part of the
signal is the same in the GTR as well as in STG and $f(R)$ theories.
This can be immediately understood by writing down explicitly, the
corresponding line-elements. In the standard GTR the line element
for a GW propagating in the $z-$direction can be written down as
\cite{key-16} 
\begin{equation}
ds^{2}=dt^{2}-dz^{2}-(1+h_{+})dx^{2}-(1-h_{+})dy^{2}-2h_{\times}dxdy,\label{eq: metrica TT totale}
\end{equation}
where $h_{+}$ and $h_{\times}$ are expressed in terms of synchronous
coordinates in the transverse-traceless (TT) gauge \cite{key-16}.
If the third mode of STG is massless the line-element in the TT gauge
can be extended with the one more polarization $h_{\Phi}$ as \cite{key-10,key-12,key-14}
\begin{equation}
ds^{2}=dt^{2}-dz^{2}-(1+h_{+}+h_{\Phi})dx^{2}-(1-h_{+}+h_{\Phi})dy^{2}-2h_{\times}dxdy.\label{eq: metrica TT super totale}
\end{equation}
On the other hand, as previously stressed, STG and $f(R)$ theories
can have the third mode being massive. In that case, it is impossible
to extend the TT gauge to the third mode because of the presence of
the small mass $m$ which generates a GW longitudinal component \cite{key-12,key-13,key-14}.
Then, gauge transformations permit to find the line-element as \cite{key-12,key-13,key-14}

\begin{equation}
\begin{array}{c}
ds^{2}=dt^{2}-dz^{2}-(1+h_{+})dx^{2}-(1-h_{+})dy^{2}-2h_{\times}dxdy+\\
\\
+(1+h_{\Phi})(dt^{2}-dz^{2}-dx^{2}-dy^{2}).
\end{array}\label{eq: metrica scalaronica}
\end{equation}
Then, one sees immediately that the tensor modes in $f(R)$ theories
and STG are the same as in the GTR independently on the issue that
the third additional mode is massless or massive. In fact, setting
$h_{\Phi}=0$ in eqs. (\ref{eq: metrica TT super totale}) and (\ref{eq: metrica scalaronica}),
one sees immediately that both eqs. (\ref{eq: metrica TT super totale})
and (\ref{eq: metrica scalaronica}) reduce to eq. (\ref{eq: metrica TT totale}).
Thus, as the tensor modes in $f(R)$ theories and STG are massless
and the same as in the GTR, the analysis in \cite{key-4} does not
work for these two classes of theories. In fact, there are other constraints
on massive theories of gravity, from weak-lensing, which are stronger
than those from the recent GW detections. Similarly, there are laboratory
experiments that constraint Yukawa-deviations from Newtonian gravity,
that place much stricter bounds on $f(R)$ gravity. A common argument
is to invoke a chameleon mechanism \cite{key-22} that screens the
deviations on certain scales. If screening is invoked then one could
argue that any constraints obtained apply only to this system or only
to BH binaries. In any case, the constraint here is coming from the
propagation of the GWs over cosmological distances rather than from
processes occurring on the scale of the binary, so it is not a local
constraint. Cosmological GWs can also put constrains on the inflaton
field \cite{key-27,key-28}. A further clarification is needed. Following
eqs. (7) - (9) one argues that in GTR, $f(R)$ gravity and STG theories
the tensor modes are the same and hence LIGO can not distinguish between
them. While it is true that the detector responds in the same way
to these modes in all theories, the evolution of the modes themselves
may not be the same. In STG there may be additional channels into
which energy is radiated as GWs. In the above discussion we identified
the third massless scalar mode. While it is right that the detector
cannot distinguish a monochromatic scalar mode signal from one in
the tensor modes, if such additional modes exist they will cause the
binary system to inspiral more quickly. This more rapid inspiral will
be visible in the phase evolution of the tensor modes and so LIGO
can still place constraints on the existence of such modes even if
they are not directly observed. 

We know that STG can be massless \cite{key-10,key-12,key-14}. Thus,
let us see what happens in the case of massless $f(R)$ theories,
that, to our knowledge, has not been analysed in the literature. In
order to linearize the $f(R)$ theories one uses the identifications
\cite{key-11} 
\begin{equation}
\begin{array}{ccccc}
\Phi\rightarrow\frac{df(R)}{dR} &  & \textrm{and } &  & \frac{dV}{d\Phi}\rightarrow\frac{2f(R)-R\frac{df(R)}{dR}}{3}.\end{array}\label{eq: identifica}
\end{equation}
The mass of the extra polarisation mode is given by \cite{key-11}
\begin{equation}
\frac{dV}{d\Phi}\simeq m^{2}\delta\Phi,\label{eq: mass}
\end{equation}
where $\delta\Phi$ is the variation of the effective scalar field
$\Phi$ near a minimum for the effective potential $V$, see \cite{key-11}
for details. Thus, for $m=0$ one gets 
\begin{equation}
2f(R)=R\frac{df(R)}{dR}.\label{eq: separa}
\end{equation}
By separating the variables eq. (\ref{eq: separa}) is easily solved
as 
\begin{equation}
f(R)=R^{2}.\label{eq: R quadro}
\end{equation}
In the general case of $f(R)$ theories, to first order in $h_{\mu\nu}$
and $\delta\Phi$, calling, $\widetilde{R}_{\mu\nu}$ and $\widetilde{R}$
the linearized quantity which correspond to $R_{\mu\nu}$ and $R$,
(where $R_{\mu\nu}$ and $R$ are the Ricci tensor and the Ricci scalar
respectively) the linearized field equations are \cite{key-11} 
\begin{equation}
\begin{array}{c}
\widetilde{R}_{\mu\nu}-\frac{\widetilde{R}}{2}\eta_{\mu\nu}=(\partial_{\mu}\partial_{\nu}h_{\Phi}-\eta_{\mu\nu}\square h_{\Phi})\\
\\
\square h_{\Phi}=m^{2}h_{\Phi}.
\end{array}\label{eq: linearizzate1}
\end{equation}
For the particular case of $f(R)=R^{2}$, eqs. (\ref{eq: linearizzate1})
become 
\begin{equation}
\begin{array}{c}
\widetilde{R}_{\mu\nu}-\frac{\widetilde{R}}{2}\eta_{\mu\nu}=(\partial_{\mu}\partial_{\nu}h_{\Phi}-\eta_{\mu\nu}\square h_{\Phi})\\
\\
\square h_{\Phi}=0.
\end{array}\label{eq: linearizzate 2}
\end{equation}
This case is the exact analogous of STG having a third massless mode
that has been discussed in detail in \cite{key-10}, see eqs. (23)
of \cite{key-10}. Thus, following the analysis in \cite{key-10}
step by step one arrives to the line element for the third component
of the GW \cite{key-10} 
\begin{equation}
ds^{2}=dt^{2}-dz^{2}-\left(1+h_{\Phi}\right)\left(dx^{2}+dy^{2}\right),\label{eq: metrica puramente scalare}
\end{equation}
that is the part of eq. (\ref{eq: metrica TT super totale}) arising
from the additional GW mode $h_{\Phi}$. Thus, from the mathematical
point of view, also $f(R)$ theories admit a third massless GW polarization.
But we recall that the class of $\alpha R^{n}$ theories (where $n$
is not restricted to be an integer and $\alpha>0$ has the dimensions
of a mass squared \cite{key-17}), is viable only for $n=1+\varepsilon$
with $0\leq\varepsilon\ll1$ \cite{key-15,key-16,key-17}. Consequently,
since the $R^{2}$ theory is not viable, we've discovered \uline{a
second, interesting new result}: the only $f(R)$ theory having a
third massless GW mode is ruled out by our previous analysis. Thus,
the extra polarization mode in $f(R)$ theories must be \emph{always
massive}. This also means that the only massless $f(R)$ theory which
results viable is the GTR, for which it is $f(R)=R$. This result
has an important consequence on the debate on the equivalence or non-equivalence
between $f(R)$ theories an STG \cite{key-7,key-12}. In fact, despite
it is well known that there is a general conformal equivalence between
STG and$f(R)$ theories, there is a big debate on the possibility
that such a conformal equivalence should be a \emph{physical equivalence}
too, see \cite{key-7,key-9,key-12,key-32}. Clearly, our result implies
indeed that these two classes of theories have \uline{only} a conformal
equivalence because, differently from $f(R)$ theories, STG admit
a third massless polarization mode. 

\section{Conclusion remarks}

The GTR is not yet ultimately confirmed by the results of the LIGO
Scientific Collaboration and the Virgo Collaborations in \cite{key-2,key-4},
{[}33 - 37{]}. In fact, on one hand, in principle, there is still
room for AGTs having massive tensor polarizations with a graviton
mass $m_{g}\leq7.7\times10^{-23}\frac{eV}{c^{2}}$ . On the other
hand, there is still room for $f(R)$ theories and STG. In fact, the
recent GW detections did not put constraints on the these two classes
of theories. Thus, we understand which is the key point here. Only
a perfect knowledge of the motion of the interferometer's mirror will
permit one to determine if the GTR is the definitive theory of gravity.
In order to ultimately conclude that the GTR is the definitive theory
of gravity, one must prove that the oscillations of the interferometer's
mirror are in fact governed by eqs. (\ref{eq: traditional GTR}).
Otherwise, if one proves that the oscillations of the interferometer's
mirror are in fact governed by eqs. (\ref{eq: massless STG}) or eqs.
(\ref{eq: massive polarization}), then the GTR must be extended.
In this framework also the results of this paper on the frequency
and angular dependent response functions (\ref{eq: risposta totale 2})
and (\ref{eq: risposta totale Virgo}) could be useful, because they
can permit to discriminate between massless and massive modes in STG
and $f(R)$ theories.

On the other hand, at the present time, the sensitivity of the current
ground based GW interferometers is not sufficiently high to determine
if the oscillations of the interferometer's mirror are governed by
eqs. (\ref{eq: traditional GTR}), or if they are governed by eqs.
(\ref{eq: massless STG}) or eqs. (\ref{eq: massive polarization}).
That sensitivity is also not sufficiently high to determine the frequency
and angular dependent response functions (\ref{eq: risposta totale 2})
and (\ref{eq: risposta totale Virgo}). A network including interferometers
with different orientations is indeed required and we're hoping that
future advancements in ground-based projects and space-based projects
will have a sufficiently high sensitivity. Such advancements would
enable physicists to determine, with absolute precision, the direction
of GW propagation and the motion of the various involved mirrors.
In other words, in the nascent GW astronomy we hope not only to obtain
new, precious astrophysical information, but we also hope to be able
to discriminate between eqs. (\ref{eq: traditional GTR}), eqs. (\ref{eq: massless STG}),
and eqs. (\ref{eq: massive polarization}) and also to discriminate
between the frequency and angular dependent response functions (\ref{eq: risposta totale 2})
and (\ref{eq: risposta totale Virgo}). Such advances in GW technology
would equip us with the means and results to ultimately confirm the
GTR or, alternatively, to ultimately clarify that the GTR must be
extended.

Summarizing, in this paper we have discussed the future of gravitational
theories in the framework of GW astronomy. In particular, we performed
a calculation of the frequency and angular dependent response function
that a GW detector would see if massive modes were present, allowing
for sources incident from any direction on the sky. In addition, we
have shown that massive (in terms of the third additional polarization)
$f(R)$ theories of gravity and STG are still alive, while there is
no room for $f(R)$ theories of gravity having a massless extra polarization
mode. 

\section{Acknowledgements }

It is a pleasure to thank my student Nathan O. Schmidt for editing
the English language of this paper. This work has been supported financially
by the Research Institute for Astronomy and Astrophysics of Maragha
(RIAAM), Research Project No. 1/4717-110.

\end{document}